\documentclass[twocolumn,showpacs,preprintnumbers,amsmath,amssymb]{revtex4}
\usepackage{graphicx}
\begin{document}
\title {High resolution superconducting single flux quantum comparator\\ for sub kelvin temperatures}

\author{A.M. Savin}
\author{J.P. Pekola}
\author{T. Holmqvist }
\affiliation{Low Temperature Laboratory, Helsinki University of Technology, P.O. Box 3500, FIN-02015 HUT, 
Finland}
\author{J. Hassel}
\author{L. Gr\"{o}nberg}
\author{P. Helist\"{o}}
\affiliation{VTT, P.O. Box 1000, 02044 VTT, Finland}
\author{A. Kidiyarova-Shevchenko}
\affiliation{Microtechnology and Nanoscience Department, Chalmers
University of Technology, 412 96 Gothenburg, Sweden }

\begin{abstract}
A design of subkelvin single flux quantum (SFQ) circuits with
reduced power dissipation and additional cooling of shunt
resistors for superconducting qubit control circuits has been
developed and characterized. We demonstrate operation of SFQ
comparators with current resolution of 40 nA at 2 GHz sampling
rate. Due to improved cooling the electron temperature in shunt
resistors of a SFQ comparator is below 50 mK when the bath
temperature is about 30 mK.

\end{abstract}


\maketitle

    Fast development of quantum information processing and its particular realization
based on superconducting quantum bits (qubits) give rise to the
need of development of fast low noise support electronics. Fully
scalable solutions require the control circuits to be placed near
qubits, i.e., at temperature far below 1 K, which significantly
limits power dissipation of the control electronics. Nowadays
single flux quantum (SFQ) digital circuits \cite{Likharev1991} are
considered as promising classical devices for qubit manipulation
and read out: they are very fast, operate at low temperature and
their power dissipation is substantially smaller as compared to
semiconductor devices. In spite of rather low power dissipation
and high sensitivity, SFQ logic requires significant modification
for operation at sub kelvin temperatures. The direct rescaling of
the existing SFQ circuits can be realized by reduction of the
operation frequency and critical currents of the junctions
\cite{Savin2006, Intiso2006}. The final choice of SFQ circuit
parameters is determined by a compromise between required speed
and possible level of power dissipation. The commercial
technological standards for niobium trilayer fabrication process
set additional limitations to the circuit parameters.

    Josephson junction (JJ) based balanced comparator
\cite{SensitivityBC, SignalResolution, BC} is a basic element of
the family of SFQ digital electronics. It represents a very fast
and accurate readout circuit for qubits and it can be used for
testing new approaches and solutions aimed at developing new
millikelvin SFQ circuits. In the regime where thermal fluctuations
dominate over quantum effects strong dependence  on temperature of
the comparator uncertainty zone (gray zone)
\cite{SignalResolution, BC, Savin2006} can be utilized for
measurements of the electronic temperature in SFQ circuits.
Generally, sensitivity requirements and SFQ circuit realizations
are different for different types of superconductor qubits, but
the width of the gray zone directly characterizes current (or
magnetic flux) sensitivity and noise properties of SFQ circuits.
In this letter we report on the results of our approach to
fabricate SFQ circuits with special attention to cooling of
resistors, and demonstrate operation of SFQ comparators with high
current resolution and low power dissipation at bath temperatures
down to 30 mK.

A few important issues should be solved to satisfy the
requirements for operation of SFQ circuit at bath temperatures
below 100 mK. Main dissipation elements in SFQ circuits are bias
and shunt resistors. The noise characteristics of the SFQ circuits
are mainly determined by shunt resistors and the effect of power
dissipation in bias resistors can be reduced by keeping them on a
separate chip. We focus our attention to the influence of the
shunt resistors and junction parameters which significantly affect
operation of SFQ circuits at low temperature. The energy
dissipated in a shunt resistor during one switching event is
$Q\approx I_{c}\Phi _{0}$, where $I_{c}$ is a critical current of
the junction and $\Phi _{0}=\frac{h}{2e}$ is the flux quantum. The
power dissipation is then proportional to the operation frequency
$f$: $P\approx f I_{c}\Phi _{0}$. As it follows from this equation
reduction of both the critical current and operation frequency
leads to lower dissipation in SFQ circuits. There are reasons why
one should not lower these values indefinitely. High operation
speed being the main advantage of SFQ logics allows one to process
a reasonable number of operations during the coherent evolution of
a qubit. From this point of view operation frequency should be
several GHz or higher. Maximum operation frequency of JJ based
logic should be well below plasma frequency of the junction, which
in the case of niobium trilayer process requires the critical
current density to be above 10 A/cm$^{2}$ for $f$ = 2 GHz. This
level is in line with the technological limits: 30 A/cm$^{2}$ is
the lowest commercially available current density. In the
situation of fixed critical current density, reduction of the JJ
critical current is limited by lithography resolution and,
finally, for the 30 A/cm$^{2}$ process, reasonable JJ critical
current is of order of few microamperes. As a result the power
dissipation in shunt resistor (assuming $I_{c}$ = 2 $\mu$A and $f$
= 2 GHz) can be reduced to 8 pW.

\begin{figure}[t!]
\begin{center}
\includegraphics[width=8.5cm,clip]{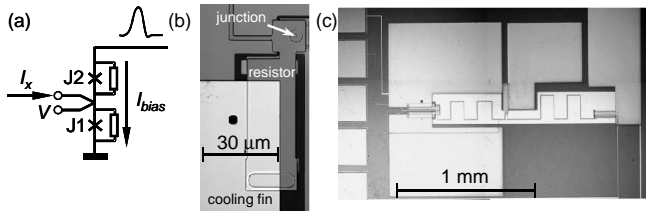}
\end{center}
\caption{Equivalent circuit of a SFQ comparator (a). Optical
photograph of a shunt resistor with connection to a cooling fin in
(b), and of the measured comparator with cooling fins in (c).}
\label{Fig1}
\end{figure}

The main bottleneck of electron cooling in the resistor is
electron-phonon coupling which is relatively weak at sub kelvin
temperatures. Electron temperature $T_{e}$ and the lattice
temperature $T_p$ in the resistor are related as follows
\cite{roukes85, wellstood}: $P_{e-p}=\Sigma \Lambda
(T_{e}^{5}-T_{p}^{5})$. Here $P_{e-p}$ is the heat flux between
the electrons and the lattice, $\Sigma$ is a material constant,
and $\Lambda$ is the volume of the resistor. For metals, typical
value of $\Sigma$ is $\Sigma \simeq 1 \cdot 10^{9}$
Wm$^{-3}$K$^{-5}$ \cite{Giazotto}. This indicates that
electron-phonon heat transport depends on the resistor volume. It
was demonstrated recently that in the case of standard resistor
size ($\Lambda\approx$ 10 $\mu$m$^{3}$) and power dissipation of
10 - 20 pW, electron temperature in a shunt resistor was about 500
mK, even when bath temperature was below 100 mK \cite{Savin2006}.
The volume of a shunt resistor realized as a thin resistive film
(with thickness 40 - 100 nm) can be increased only up to 100 - 500
$\mu$m$^{3}$ without introducing harmful parasitic capacitance.
Further cooling of electrons can be achieved by connecting to the
resistors cooling fins with large volume and high thermal
conductivity \cite{Savin2006, Intiso2006}. Following these
requirements Nb trilayer process with additional 800 nm thick
copper layer \cite{VTTprocess} has been developed for sub kelvin
SFQ circuits. Theoretical analysis of thermal budget of circuits
with reduced critical currents and low operation frequency
predicts possibility to reduce electron temperature in shunt
rersistors to below 100 mK \cite {Intiso2006}.

\begin{figure}[t!]
\begin{center}
\includegraphics[width=8.0cm,clip]{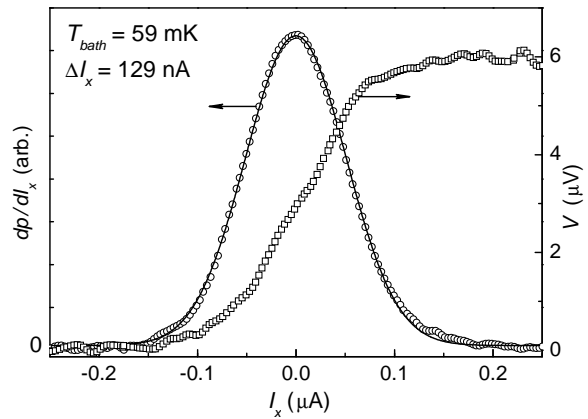}
\end{center}
\caption{Voltage across junction J1 (squares, right scale) and
density of switching probability (circles, left scale) as a
function of the signal current $I_{x}$. Solid line corresponds to
a Gaussian fit.}
 \label{GrZG}
\end{figure}

A balanced SFQ comparator \cite{SensitivityBC, SignalResolution}
consists of two identical overdamped JJs (J1 and J2 in Fig.
\ref{Fig1}a) biased in series by dc $I_{bias}$ below critical
current. SFQ pulses generated by a driver (a JJ biased by constant
voltage) are applied to the comparator. SFQ pulses escape from the
circuit either via junction J1 or via J2 depending on the sign of
the signal current $I_{x}$. Thermal and quantum fluctuations
introduce uncertainty between these two processes. Detailed
theoretical analyses of the gray zone for the balanced comparator
have been done \cite {SensitivityBC, Likharev2001}, and the
theoretical model demonstrates good agreement with experiment
\cite {SignalResolution, BC}. At relatively low SFQ pulse
repetition rate, the dependence of switching probability ($p$) of
one of the two junctions on signal current $I_{x}$ is described by
an error function and the probability density $\frac{dp}{dI_{x}}$
is represented by Gaussian distribution. Gray zone of the
comparator $\Delta I_{x}$ is usually defined as \cite
{SensitivityBC, SignalResolution, BC, Likharev2001}
\begin{equation}
\Delta I _{x} \equiv \mid \frac{dp}{dI_{x}} \mid ^{-1} _{p=1/2}.
\label{GrZDef}
\end{equation}

Thermal noise limits sensitivity of the balanced comparator at
high temperatures, at low temperature quantum noise starts to
dominate. The crossover temperature $T^*$ between thermal and
quantum limits is determined by comparator parameters: $T^* \simeq
eV_c/\pi k_{B}$, where $V_c=I_{c}R_{s}$. According to conventional
SFQ design priorities the value of shunt resistors, $R_{s}$, in
SFQ circuits corresponds to the critical damping $\beta \simeq$ 1.
Furthermore, higher resistance value is desirable for reduction of
current noise generated by shunt resistor. However, in order to
scale the gray zone down with reduction of temperature, the
comparator should be in the thermal limit at the lowest bath
temperature (or, more exactly, at the lowest electron temperature
of shunt resistors), which may require stronger damping via
reduction of resistance of shunts. The width $\Delta I_{x}$ of the
``gray zone'' in the regime dominated by thermal fluctuations
depends on JJ critical current and temperature as(see, e.g.,
\cite{SensitivityBC, Likharev2001})
\begin{equation}
\Delta I _{x} \simeq \alpha (2\pi I_{T}I_{C})^{1/2}. \label{GrZ}
\end{equation}
Here $I_{T}=2\pi k_{B}T/\Phi_{0}$ and $\alpha$ is a dimensionless
parameter of order unity determined by comparator and driver
characteristics.

\begin{figure}[b!]
\begin{center}
\includegraphics[width=8.0cm,clip]{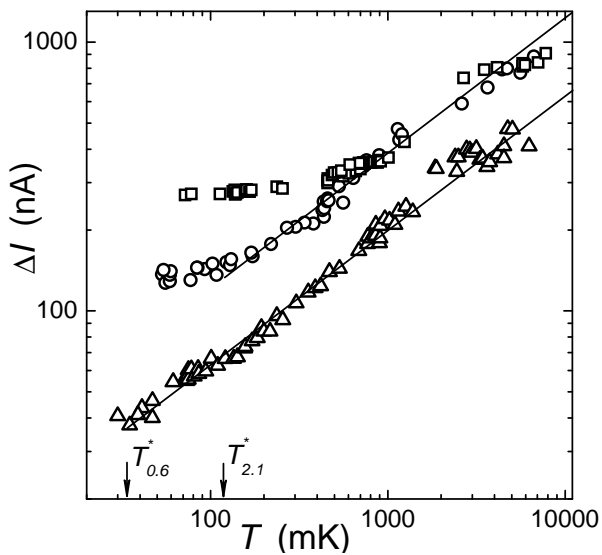}
\end{center}
\caption{Temperature dependence of the gray zone for the
comparator without cooling fins ($I_{c}$ = 2.1 $\mu$A (squares))
and for the comparators with cooling fins ($I_{c}$ = 2.1 $\mu$A
(circles), $I_{c}$ = 0.6 $\mu$A (triangles)). Solid lines
represent corresponding theoretical dependencies of the gray zone
in thermal limit (Eq. \ref{GrZ} with $\alpha$ = 0.5). Crossover
temperatures between thermal and quantum limits for the
comparators with $I_{c}$ = 2.1 $\mu$A ($T_{2.1}^* $) and $I_{c}$ =
0.6 $\mu$A ($T_{0.6}^* $) are marked by arrows.}
 \label{GrZCoolF}
\end{figure}

The comparators (Fig.~\ref{Fig1}) were fabricated using 30
A/cm$^{2}$ Nb trilayer process with additional 800 nm thick Cu
layer used for heat sinks of shunt resistors and with following
parameters: $I_{c}$ = 2.1 $\mu$A, $R_{s}$ = 15.4 $\Omega$. For
these parameters, the junctions are overdamped, and the crossover
temperature $T^*$ between the regimes of thermal and quantum
fluctuations is 120 mK. The volume of the cooling fins is about 4
x 10$^5 \mu$m$^3$. The comparator with similar parameters but
without cooling fins was fabricated on the same chip in order to
compare operation of the two devices. Figures ~\ref{Fig1}b and
~\ref{Fig1}c illustrate practical implementation of cooling fins
in SFQ circuits.

The measurement procedure is based on the idea suggested in
Ref.~\cite{BC}: the width $\Delta I_{x}$ of the gray zone can be
obtained by the measurements of dc voltage $V$ across one of the
comparator junctions as a function of the applied current $I_{x}$.
To increase the accuracy of the gray zone measurements we perform
modulation measurements with lock-in technique: the low frequency
(5-20 Hz) modulation of $I_{x}$ is used and corresponding low
frequency ac voltage across the comparator junction is measured.
This way we measure directly the probability density
$\frac{dp}{dI_{x}}$. The measurements were carried out at the
frequency range essentially below plasma frequency where the
dependence of $\frac{dp}{dI_{x}}$ on $I_{x}$ does not deviate
noticeably from Gaussian. Typical measured dc voltage across
comparator junction J1 and probability distribution for the
comparator with cooling fins at bath temperature 59 mK are
presented in Fig.~\ref{GrZG}. The value of the gray zone was
derived from a Gaussian fit of experimental data (solid line in
Fig.~\ref{GrZG}). The difference between gray zone values
calculated from fitting results and directly from experimental
data using Eq. \ref{GrZDef} was below 5\% for the data presented
below. At higher operation frequencies (above 5 GHz) measured
probability distribution starts to deviate from Gaussian form
which was used to identify the upper operation frequency of the
comparator.

Temperature dependence of gray zone for the comparators without
and with cooling fins  operating at 3 GHz are presented in
Fig.~\ref{GrZCoolF}. The solid line corresponds to the theoretical
prediction for thermal limit [Eq.~(\ref{GrZ})] assuming that the
effective electron temperature of the resistors coincides with the
bath temperature $T$. The saturation temperatures are very
different for the two comparators. Whereas the gray zone of the
comparator without cooling fins saturates at the value
corresponding to $T_{e}\simeq$ 400 mK, the comparator with cooling
fins demonstrates further reduction of the gray zone with
saturation temperature well below 200 mK. Observed saturation of
the gray zone for the comparator without cooling fins is clearly
above quantum limit and caused by overheating of electrons. The
saturation temperature for the comparator with cooling fins is
only slightly higher than the crossover temperature $T^*$ = 120 mK
for this comparator, and in this case the quantum fluctuations
seem to affect the final width of the gray zone.

Subsequent increase of the comparator resolution can be achieved
by further reduction of the critical current. The optimal way to
achieve this without degradation of frequency characteristics of
the comparator is reduction of junction size. Due to the
resolution of lithography we reduced critical current by reduction
of the critical current density down to 10 A/cm$^{2}$. In this
case the maximum operation frequency of a balanced comparator was
only slightly above 2 GHz, but it was enough to demonstrate
increase of sensitivity. The temperature dependence of the gray
zone for the comparator with cooling fins and reduced critical
current ($I_{c}$ = 0.6 $\mu$A, $R_{s}$ = 15.4 $\Omega$) is
represented in Fig. \ref{GrZCoolF}. The increase of the comparator
resolution is obtained due to both lower critical current and
lower power dissipation. High current resolution and low
equivalent noise temperature achieved confirm the good
perspectives of SFQ circuits as control and read out devices for
superconducting qubits.

In summary, we presented an approach for design of subkelvin SFQ
circuits with reduced power dissipation for superconducting
qubits. We have fabricated and tested a basic SFQ circuit - a
balanced comparator - at bath temperatures down to 30 mK. Due to
reduced power dissipation and improved cooling of the shunt
resistors the effective noise temperature of SFQ devices can be
reduced to below 50 mK, and current resolution of SFQ comparators
can be as high as 40 nA at 2 GHz sampling rate.

This work was supported by Academy of Finland and "RSFQubit" FP6
project of European Union.


\end{document}